\documentclass[conference]{IEEEtran}
\usepackage{graphicx}
\usepackage{subfigure}
\usepackage{color}
\usepackage{float}
\usepackage{amsmath,amssymb,amsfonts,amsthm}
\usepackage{chngpage}
\usepackage{array}
\usepackage{algorithm}
\usepackage{algpseudocode}
\usepackage{amssymb}
\usepackage{tabulary}
\usepackage{booktabs}
\usepackage{cases}
\usepackage{geometry}

\newtheorem{thm} {Theorem}

\newtheorem{cor} {Corollary}

\makeatletter
\floatname{algorithm}{Procedure}

\newcommand{\Rmnum}[1]{\expandafter\@slowromancap\romannumeral #1@}
\makeatother
\input{tcilatex}
\begin{document}

\title{Multicast Group Management for Multi-View 3D Videos in Wireless Networks}
\author{\IEEEauthorblockN{Chi-Heng Lin$^1$,~De-Nian Yang$^1$,~Chih-Chung Lin$^1$ and~Wanjiun Liao$^2$}\
\IEEEauthorblockA{Academia Sinica, Taipei, Taiwan$^1$\\
\{cchen2008, dnyang, chchlin\}@iis.sinica.edu.tw$^1$\\
National Taiwan University, Taipei, Taiwan$^2$\\
wjliao@cc.ee.ntu.edu.tw$^2$}}

\maketitle

\begin{abstract}
With the emergence of naked-eye 3D mobile devices, mobile 3D video services
become increasingly important for video service providers, such as Youtube
and Netflix, while multi-view 3D videos are potential to bring out varied
innovative applications. However, enabling multi-view 3D video services may
overwhelm WiFi networks when we multicast every view of a video. In this
paper, therefore, we propose to incorporate depth-image-based rendering
(DIBR), which allows each mobile client to synthesize the desired view from
nearby left and right views, to effectively reduce the bandwidth
consumption. Moreover, due to varied channel conditions, each client may
suffer from different packet loss probabilities, and retransmissions incur
additional bandwidth consumption. To address this issue, we first analyze
the merit of view protection via DIBR for multi-view video multicast and
then design a new protocol, named Multi-View Group Management Protocol
(MVGMP), for the dynamic group management of multicast users. Simulation
results manifest that our protocol effectively reduces the bandwidth
consumption and increases the probability for each client to successfully
playback the desired view of a multi-view 3D video.
\end{abstract}


%\author{\IEEEauthorblockN{None, None, None}
%\IEEEauthorblockA{
%Email: dnyang@iis.sinica.edu.tw} }


\section{Introduction}

The \IEEEPARstart{I}{EEE} 802.11 \cite{Stand2012} WiFi standard has achieved
massive market penetration due to its low cost, easy deployment and high
bandwidth. Also, with the recent emergence of naked-eye 3D mobile devices,
such as Amazon's 3D Fire Phone, HTC's EVO 3D, LG's Optimus 3D, and Sharp's
Lynx, mobile 3D video services are expected to become increasingly important
for video service providers, such as Youtube and Netflix. In contrast to
traditional stereo single-view 3D video formats, multi-view 3D videos
provide users with a choice of viewing angles and thus are expected to
inspire the development of innovative applications in television, movies,
education, and advertising \cite{Signal2009}.
Previous researches on the deployment of 3D videos in wireless networks
mostly focused on improving 3D video quality for single-view 3D videos \cite%
{ICNC2005, ICC2013, ICCF2013}. Nevertheless, multi-view 3D videos, which
typically offer 16 different viewing angels \cite{Meet2008}, are expected to
significantly increase the network load when all views are transmitted.

One promising way to remedy the bandwidth issue is by exploiting
depth-image-based rendering (DIBR) in mobile clients. Because adjacent views
usually share many similar parts, a user's desired view can be synthesized
from nearby left and right views \cite{Signal2009}. Several schemes for bit
allocation between the texture and depth map \cite{Broadcasting2011,
Multimedia2012} and rate control with layered encoding for a multi-view 3D
video \cite{CISP2011, ACMMultimedia2012} have been proposed to ensure that
the quality of the synthesized view is very close to the original view
(i.e., by minimizing total distortion or maximizing quality). Therefore,
exploiting DIBR in clients eliminates the requirement to deliver each view
of a multi-view video if the left and right views of the desired view have
been transmitted to other clients. Moreover, the computation overhead
incurred by DIBR is small enough to be handled by current mobile devices
\cite{ACMMultimedia2012, Processing2012}.

However, multi-view 3D video multicast with DIBR brings new challenges in
WiFi networks. 1) The number of views between the left and right transmitted
views needs to be constrained to ensure the quality of the synthesized view
\cite{Signal2009}. In other words, since each transmitted view is shared by
multiple clients, one must carefully select the transmitted views so that
the desired view of each user can be synthesized with good quality. DIBR has
a quality constraint \cite{Signal2009}, which specifies that the left and
right views are allowed to be at most $R$ views away (i.e., $R-1$ views
between them) to ensure that every view between the left and right view can
be successfully synthesized with good quality. Therefore, each new user
cannot arbitrarily choose a left and a right view for synthesis with DIBR.
2) WiFi networks frequently suffer from wireless erasure, and different
clients suffer from different loss probabilities due to varying channel
conditions \cite{ICC2005, INFOCOM2007, IWCMC2014, IICSP2007}. In 2D and
single-view 3D videos, the \textit{view loss probability} for each user can
be easily derived according to the corresponding channel state information.
The view loss probability for each user is correlated to the selected
bit-rate, channel, and the setting of MIMO (ex. antennas, spatial streams)
in 802.11 networks. For multi-view 3D videos, however, when a video frame is
lost for a user $i$ subscribing a view $k_{i}$, the left and right views
multicasted in the network for other users can natively serve to \textit{%
protect} view $k_{i}$, since the user $i$ can synthesize the desired view
from the two views using DIBR. However, the view synthesis will fail if only
one left view or one right view is received successfully by the client.
Therefore, a new research problem is to find out the \textit{view failure
probability}, which is the probability that each user doesnot successfully
receive and synthesize his/her desired view.

In this paper, therefore, we first analyze the merits of DIBR for multi-view
3D video multicast in multi-rate multi-channel WiFi networks \cite{Stand2012}.
We analyze the view failure probability for comparison with the traditional
view loss probability. We then propose Multi-View Group Management Protocol
(MVGMP) for multi-view 3D multicast. When a user joins the video
multicast group, it can exploit our analytical results to request the AP to
transmit the most suitable right and left views, so that the view failure
probability is guaranteed to stay below a threshold. On the other hands,
when a user leaves the video multicast group, the proposed protocol
carefully selects and withdraws a set of delivered views to reduce the
network load, so that the video failure probability for other users will not
exceed the threshold. Bandwidth consumption can be effectively reduced since
not all subscribed views are necessary to be delivered. Moreover, the
protocol supports the scenario in which each user subscribes to multiple
desired views.

The rest of the paper is organized as follows. Section II describes the
system model. Section III analyzes the view loss probability and view
failure probabilities. Section IV presents the proposed protocol. Section V
shows the simulation results, and Section VI concludes this paper.

\section{System Model}

This paper considers the single-cell point-to-multipoint video transmissions
in IEEE 802.11 networks, where the views transmitted by different bit-rates
and on different channels are associated with different loss probabilities
\cite{AdHoc2011, ICC2005, INFOCOM2007, IWCMC2014, IICSP2007}. Currently,
many video services, such Youtube and Netflix, require reliable
transmissions since Flash or MPEG DASH \cite{Dash20} are exploited for video streaming. Nevertheless, the
current IEEE 802.2 LLC protocol for IEEE 802.11 networks does not support
reliable multicast transmissions \cite{Stand1998}, and error recovery
thereby needs to be handled by Layer-3 reliable multicast standards, such as
PGM \cite{PGM2001}.

A 3D video in multi-view plus depth can be encoded by varied encoding
schemes \cite{Circuits2007, Proceed2011}. The idea of DIBR is to infer and
synthesize the parts different from nearby views, while effective techniques
are proposed to ensure the video quality \cite{Processing2011, Multi2011}.
In the original WiFi multicast without DIBR, AP separately multicasts each
view to the clients. By contrast, a desired view can be synthesized be
nearby left and right views with DIBR, while the quality constraint in DIBR
states that there are at most $R-1$ views between the left and right views,
and $R$ can be set according to \cite{Signal2009}. In addition, when the
subscribed view is lost for a user, the user can also try to synthesize the
view according to the left view and right view. In the next section, we
investigate the merits of DIBR by analyzing the view failure probability for
comparison with the traditional view loss probability.

\section{Analytical Solution\label{sec: analysis}}

In this section, we present the analytical results for multi-rate
multi-channel IEEE 802.11 networks with DIBR. We first study the scenario of
single-view subscription for each client and then extend it to multi-view
subscription. Table \uppercase\expandafter{\romannumeral1} summarizes the
notations in the analysis. Based on the mathematical analysis, a new
protocol is proposed in the next section to adaptively assign the proper
views to each client.

\begin{table}[t]
\caption{Notations.}
\label{table1}\vspace{-12pt}
\par
\begin{center}
\begin{tabular}{|l|l|}
\hline
\textbf{Description} & \textbf{Notation} \\ \hline
$R$ & Quality constraint of DIBR \\ \hline
$M$ & Total number of views which can be selected \\ \hline
$k_i$ & The $k$ view selected by user $i$ \\ \hline
$D_i$ & A set of the available data rates for user $i$ \\ \hline
$C_i$ & A set of the available channels for user $i$ \\ \hline
$n_{j,c,r}$ & Number of broadcasts for the $j$ view transmitted \\
& by the $r$ rate in the $c$ channel within a reasonable \\
& frame time $T_{f}$ \\ \hline
$p_{i,c,r}$ & The loss probability for user $i$ under the $c$ channel \\
& and the $r$ rate \\ \hline
$P_{\varepsilon}^{(i)}$ & The probability that user $i$ cannot obtain his/her
\\
& selected view either by direct reception or by DIBR \\ \hline
$p_{c,r}^{\text{AP}}(n)$ & The probability that AP broadcasts a view by $n$
times \\
& under the channel $c$ and the rate $r$ \\ \hline
$\alpha_i$ & The probability of the selected view obtained by user $i$ \\
\hline
$\eta_i$ & Minimum retransmission rate for user $i$ \\ \hline
$p_{\text{select}}$ & The probability that an user selects a certain view \\
\hline
\end{tabular}%
\end{center}
\par
\vspace{-12pt}
\end{table}

\subsection{Single View Subscription}

In single-view subscription, each user $i$ specifies only one desired view $%
k_{i}$. Each view can be sent once or multiple times by the AP. Let $%
p_{i,c,r}$ represent the \textit{view loss probability}\footnote{%
Many data frames in layer-2 aggregate one view. Thus, the loss probability
of each loss frame forms the view loss probability.}, which is the
probability that user $i$ does not successfully receive a view under the
channel $c$ and the data rate $r$. We define a new probability $%
P_{\varepsilon }^{(i)}$ for multi-view 3D videos, called \textit{view
failure probability}, which is the probability that user $i$ fails to
receive or synthesize his desired view because the desired view and nearby
left and right views for synthesis with DIBR are all lost. In other words,
the view loss probability considers only one view, while the view failure
probability jointly examines the loss events of multiple views.

\begin{thm}
For single-view subscription, the view failure probability for user $i$ is
\begin{align}
& P_{\varepsilon }^{(i)}=\prod_{c\in C_{i},r\in
D_i}p_{i,c,r}^{n_{k_{i},c,r}}\times  \notag  \label{formula1} \\
& \Bigg[\mathbf{1}\{k_{i}=1\}+\mathbf{1}\{k_{i}=M\}+  \notag \\
& \sum_{k=1}^{R-1}\Bigg((1-\prod_{c^{\prime }\in C_{i},r^{\prime }\in
D_i}p_{i,c^{\prime },r^{\prime }}^{n_{k_{i}-k,c^{\prime },r^{\prime
}}})\prod_{l=1}^{\min (R-k,M-k_{i})}\prod_{\underset{c_{1},c_{2}\in C_{i}}{%
r_{1},r_{2}\in D_i}}  \notag \\
&
\prod_{q=1}^{k-1}p_{i,c_{1},r_{1}}^{n_{k_{i}-q,c_{1},r_{1}}}p_{i,c_{2},r_{2}}^{n_{k_{i}+l,c_{2},r_{2}}}%
\mathbf{1}\{M-1\geq k_{i}\geq k+1\}\Bigg)+  \notag \\
&\prod_{q=1}^{\min (R-1,k_{i}-1)}\prod_{c_{3}\in C_{i},r_{3}\in
D_i}p_{i,c_{3},r_{3}}^{n_{k_{i}-q,c_{3},r_{3}}}\mathbf{1}\{M-1\geq k_{i}\geq
2\}\Bigg]  \notag
\end{align}%
where $\mathbf{1}\{\cdot \}$ denotes the indicator function.
\end{thm}

\textit{Proof:} The view failure event occurs when one of the following two
conditions holds: 1) user $i$ does not successfully receive his/her desired
view, and 2) user $i$ fails to receive any feasible set of a left view and a
right view with the view distance at most $R$ to synthesize the desired
view. The probability of the first condition is $\prod_{c\in C_{i},r\in
D_i}p_{i,c,r}^{n_{k_{i},c,r}}$ when the the desired view $k_{i} $ of user $i$
is transmitted by $n_{k_{i}}$ times. Note that if the desired view of user $i
$ is view $1$ or view $M$, i.e., $k_{i}=1$ or $k_{i}=M $, user $i$ is not
able to synthesize the desired view with DIBR, and thus the view failure
probability can be directly specified by the first condition. For every
other user $i$ with $M-1\geq k_{i}\geq 2$, we define a set of
non-overlapping events $\{\mathcal{B}_{k}\}_{k=0}^{R-1}$, where $\mathcal{B}%
_{k}$ with $k>0$ is the event that the nearest left view received by user $i$
is $k_{i}-k$ , but user $i$ fails to receive a feasible right view to
synthesize the desired view. On the other hand, $\mathcal{B}_{0}$ is the
event that the user $i$ fails to receive any left view. Therefore, $%
\bigcup_{k=0}^{R-1}\mathcal{B}_{k}$ jointly describes all events for the
second condition.

Since the two events described above are independent, we can derive their
probabilities separately and then multiply them together to obtain the view
failure probability for the user $i$.

For each event $\mathcal{B}_{k}$ with $k>0$,
\begin{align}
& P(\mathcal{B}_{k})=(1-\prod_{c^{\prime }\in C_{i},r^{\prime }\in
D_i}p_{i,c^{\prime },r^{\prime }}^{n_{k_{i}-k,c^{\prime },r^{\prime
}}})\prod_{l=1}^{\min (R-k,M-k_{i})}  \notag \\
&\prod_{\underset{c_{1},c_{2}\in C_{i}}{r_{1},r_{2}\in D_i}%
}\prod_{q=1}^{k-1}p_{i,c_{1},r_{1}}^{n_{k_{i}-q,c_{1},r_{1}}}p_{i,c_{2},r_{2}}^{n_{k_{i}+l,c_{2},r_{2}}}%
\mathbf{1}\{M-1\geq k_{i}\geq k+1\}  \notag
\end{align}%
The first term $1-\prod_{c^{\prime }\in C_{i},r^{\prime }\in
D_i}p_{i,c^{\prime },r^{\prime }}^{n_{k_{i}-k,c^{\prime },r^{\prime }}}$
represents that user $i$ successfully received view $k_{i}-k $, and the
second term
\begin{equation*}
\prod_{l=1}^{\min (R-k,M-k_{i})}\prod_{\underset{c_{1},c_{2}\in C_{i}}{%
r_{1},r_{2}\in D_i}%
}\prod_{q=1}^{k-1}p_{i,c_{1},r_{1}}^{n_{k_{i}-q,c_{1},r_{1}}}p_{i,c_{2},r_{2}}^{n_{k_{i}+l,c_{2},r_{2}}}
\end{equation*}%
describes that user $i$ does not successfully receive any left view between $%
k_{i}-k$ and $k$ and any right view from $k_{i}+1$ to $k_{i}+\min
(R-k,M-k_{i})$. It is necessary to include a indicator function in the last
term since $\mathcal{B}_{k}$ will be a null event if $k_{i}\leq k$, i.e.,
user $i$ successfully receives a view outside the view boundary. Finally,
the event $\mathcal{B}_{0}$ occurs when no left view successfully received
by user $i$.
\begin{align}
&P(\mathcal{B}_{0})=  \notag \\
&\prod_{q=1}^{\min (R-1,k_{i}-1)}\prod_{c_{3}\in C_{i},r_{3}\in
D_i}p_{i,c_{3},r_{3}}^{n_{k_{i}-q,c_{3},r_{3}}}\mathbf{1}\{M-1\geq k_{i}\geq
2\}  \notag
\end{align}%
The theorem follows after summarizing all events. $\blacksquare $ \newline

\textbf{Remark:} The merit of multi-view 3D multicast with DIBR can be
unveiled by comparing the view loss probability and view failure
probability. The view failure probability attaches a new term (i.e., the
probability of $\bigcup_{k=0}^{R-1}\mathcal{B}_{k}$) with the value at most
1 to the view loss probability, while a larger $R$ leads to a smaller
probability.

\subsection{Multiple View Subscription}

In the following, we explore the case that a user desires to subscribe
multiple views. We study the following two scenarios: 1) every view is
multicasted; 2) only one view is delivered for every $\widetilde{R}$ views, $%
\widetilde{R}$ $\leq R$, and thus it is necessary for a user to synthesize
other views accordingly. We first define $\alpha _{i}$, which represents the
percentage of desired views that can be received or synthesized by user $i$
successfully.
\begin{equation*}
\alpha _{i}=\frac{\sum_{k_{i}\in \mathcal{K}_{i}}\mathbf{1}\{\text{user }i%
\text{ can obtain view }k_{i}\}}{|\mathcal{K}_{i}|}
\end{equation*}%
where $\mathcal{K}_{i}$ denotes the set of desired views for user $i$. Since
retransmission is necessary to be involved when a desired view cannot be
received or synthesized, we derive the minimal number of views required to
be retransmitted to obtain all desired views for each user later in \cite%
{CORR}.\ By using Theorem 1, we can immediately arrive at the following
corollary.

\begin{cor}
\begin{align}  \label{formula2}
\mathbb{E}[\alpha_i]=\frac{\sum_{k_i\in\mathcal{K}_i}P_{%
\varepsilon}^{(i)}(k_i)}{|\mathcal{K}_i|}
\end{align}
where $P_{\varepsilon}^{(i)}(k_i)$ is given in Theorem 1.
\end{cor}

\textit{Proof:}
\begin{align}
\mathbb{E}[\alpha _{i}]=& \frac{\sum_{k_{i}\in \mathcal{K}_{i}}\mathbb{E}%
\mathbf{1}\{\text{user }i\text{ can obtain view }k_{i}\}}{|\mathcal{K}_{i}|}
\notag \\
=& \frac{\sum_{k_{i}\in \mathcal{K}_{i}}P_{\varepsilon }^{(i)}(k_{i})}{|%
\mathcal{K}_{i}|}  \notag
\end{align}%
$\blacksquare $

Eq. (1) becomes more complicated as $|\mathcal{K}_{i}|$ increases. In the
following, therefore, we investigate asymptotic behavior $\alpha _{i}$ for a
large $|\mathcal{K}_{i}|$ and also a large $M$ since $|\mathcal{K}_{i}|\leq
M $. To find the closed-form solution, we first consider uniform view
subscription and assume that user $i$ subscribes each view $j$ with
probability $p_{\text{select}}=\frac{|\mathcal{K}_i|}{M}$ independently
across all views so that the average number of selected views is $|\mathcal{K%
}_i|$. Assume the AP multicasts view $j$ in channel $c$ with rate $r$ by $n$
times with probability $p_{j,c,r}^{\text{AP}}(n)$ independently across all
views, channels, and rates. Although a multi-view 3D videos usually contains
only dozens of views, Section V manifests that the asymptotic analysis
result is very close to the result in Theorem 2. The theoretical result is
first summarized in the following theorem where we fix $p_{\text{select}}$
and let $|\mathcal{K}_i|\rightarrow \infty $, and we then present the
insights from the theorem by comparing the results of single-view
subscription and multi-view subscription. Due to the space constraint, a
more general analysis that also allows each user to subscribe a sequence of
consecutive views is presented in \cite{CORR}.

\begin{thm}
In mutli-view subscription,
\begin{align}
\alpha _{i}(\mathcal{K}_i)\overset{a.s.}{\rightarrow }& (1-p_i)\Bigg\{%
\sum_{k=1}^{R}k(1-p_i)p_i^{k-1}+p_i^{R}\Bigg\} \\
\mathbb{E}[\alpha _{i}(\mathcal{K}_i)]\overset{a.s.}{\rightarrow }& (1-p_i)%
\Bigg\{\sum_{k=1}^{R}k(1-p_i)p_i^{k-1}+p_i^{R}\Bigg\}
\end{align}%
as $|\mathcal{K}_i|\rightarrow \infty $, where $p_i=\prod_{c\in C_{i},r\in
D_i}\sum_{n}p_{c,r}^{\text{AP}}(n)p_{i,c,r}^{n}$
\end{thm}

\textit{Proof:} We first derive the view loss probability for user $i$.
Suppose that the AP multicasts a view by $n$ times via channel $c$ and rate $%
r$. The probability that user $i$ cannot successfully receive the view is $%
p_{i,c,r}^{n}$. Because the AP will multicast a view by $n$ times via
channel $c $ and rate $r$ with probability $p_{c,r}^{\text{AP}}(n)$, the
probability that user $i$ cannot receive the view via channel $c$ and rate $r
$ is $\sum_{n}p_{c,r}^{\text{AP}}(n)p_{i,c,r}^{n}$. Therefore, the view loss
probability for user $i$ is the multiplication of the view loss
probabilities in all channels and rates, i.e., $\prod_{c\in C_{i},r\in
D_i}\sum_{n}p_{c,r}^{\text{AP}}(n)p_{i,c,r}^{n}$. For simplification, we
denote $p_i$ as the view loss probability for user $i$ in the remaining of
proof.

Since the multicast order of views is not correlated to $\alpha _{i}$, we
assume that the AP multicasts the views from view $1$ to view $M$
sequentially. Now the scenario is similar to a tossing game, where we toss $%
M $ coins, a face-up coin represents a view successfully receiving from the
AP, thus the face-up probability of coin is $1-p_i$. Now we mark a coin with
probability $p_{\text{select}}$ if it is face-up or if there are one former
tossed face-up coin and one latter tossed face-up coin with the view
distance at most $R$. Since the above analogy captures the mechanism of
direct reception and DIBR of views, the marked coins then represent the
views selected by user $i$ that can also be successfully obtained by him.

To derive the closed-form asymptoticl result, we exploit the delayed renewal
reward process, in which a cycle begins when a face-up coin appears, and the
cycle ends when the next face-up coin occurs. The reward is defined as the
total number of marked coins. Specifically, let $\{N(t):=\sup
\{n:\sum_{i=0}^{n}X_{i}\leq t\},t\geq 0\}$ denote the delayed renewal reward
process with interarrival time $X_{n}$, where $X_{n}$ with $n\geq 1$ is the
time difference between two consecutive face-up coins, and $X_{0}$ is the
time when the first face-up coin appears.

Let $R(M)$ and $R_{n}$ denote the total reward earned at the time $M$, which
correponds to the view numbers in a multi-view 3D video. At cycle $n$,
\begin{equation*}
\frac{R(M)}{M}=\frac{\sum_{n=1}^{N(M)}R_{n}}{M}+o(1)~~~a.s.
\end{equation*}%
where the $o(1)$ term comes from the fact that the difference between total
reward and $\sum_{n=1}^{N(M)}R_{n}$ will have a finite mean. Recall that the
reward earned at each cycle is the number of marked coins,
\begin{numcases}{\mathbb{E}[R_n|X_n]=}
   p_{\textrm{select}},  & for $X_n > R$\nonumber\\
   X_np_{\textrm{select}}, & for $X_n\leq R$
  \end{numcases}
since when $X_{n}\leq R$, there are $X_{n}$ coins can be marked (each marked
with probability $p_{\text{select}}$) between two consecutive face-up coins,
so the expectation of reward given $X_n$ is $X_np_{\text{select}}$. By
contrast, only one coin can be marked with probability $p_{\text{select}}$
when $X_{n}>R$, so the expectation of reward given $X_n$ is only $p_{\text{%
select}}$.

Since $X_{n}$ is a geometric random variable with parameter $p$, we have
\begin{equation*}
\mathbb{E}[X_{n}]=1-p_i+2(1-p_i)p_i^{2}+3(1-p_i)p_i^{3}+\cdots =\frac{1}{%
1-p_i}
\end{equation*}%
and
\begin{align}
\mathbb{E}[R_{n}]=& p_{\text{select}}p_i(1-p_i)+2p_{\text{select}%
}p_i(1-p_i)^{2}+\cdots  \notag \\
& +Rp_{\text{select}}(1-p_i)p_i^{R-1}+p_{\text{select}}p^{R}
\end{align}%
By theorem 3.6.1 of renewal process in \cite{ross},
\begin{align}
\frac{\sum_{n=1}^{N(M)}R_{n}}{M}& \overset{a.s.}{\rightarrow }\frac{\mathbb{E%
}R_{n}}{\mathbb{E}X_{n}}  \notag \\
& =p_{\text{select}}(1-p_i)\Bigg\{\sum_{k=1}^{R}k(1-p_i)p_i^{k-1}+p_i^{R}%
\Bigg\}
\end{align}%
Let $U_{M}$ denote the number of views selected by user $i$, we can write
\begin{equation*}
\alpha _{i}=\frac{R(M)}{U_{M}}=\frac{R(M)}{M}\frac{M}{U_{M}}
\end{equation*}%
For $\frac{U_{M}}{M}\overset{a.s.}{\rightarrow }p_{\text{select}}$, by the
strong law of large number, after combining with Eq. (4), (5), (6),
\begin{equation*}
\alpha _{i}\overset{a.s.}{\rightarrow }(1-p_i)\Bigg\{%
\sum_{k=1}^{R}k(1-p_i)p_i^{k-1}+p_i^{R}\Bigg\}
\end{equation*}%
The proof for convergence in mean is similar, it is only necessary to
replace convergence in Eq. (6) by the convergence in mean, which is
guaranteed by the same theorem. $\blacksquare $

\textbf{Remark:} Under the above uniform view subscription, we see that $%
\alpha _{i}$ is actually irrelevant to $p_{\text{select}}$ which implies
that different users with different number of subscription will obtain the
same percentage of views they select. Most importantly, $\alpha _{i}=1-p_i$
for multi-view 3D multicast without DIBR. In contrast, multi-view 3D
multicast without DIBR effectively improves $\alpha _{i}$ by $%
\sum_{k=1}^{R}k(1-p_i)p_i^{k-1}+p_i^{R}$. Since this term is strictly
monotonically increasing with $R$, we have $%
\sum_{k=1}^{R}k(1-p_i)p_i^{k-1}+p_i^{R}>
\sum_{k=1}^{1}k(1-p_i)p_i^{k-1}+p_i=1$, which implies the percentage of
obtained views is strictly larger in statistic by utilizing DIBR technique.

In the following, we consider the second case with only one view delivered
for every $\widetilde{R}$ views, where the bandwidth consumption can be
effectively reduced. Note that the following corollary is equivalent to
Theorem 2 when $\widetilde{R}=1$.

\begin{cor}
If the AP only transmits one view with probability $p_{c,r}^{\text{AP}}(n)$
for every $\widetilde{R}$ views,
\begin{align}
\alpha _{i}(\mathcal{K}_i)\overset{a.s.}{\rightarrow }& \frac{(1-p_i)\Bigg\{%
\sum_{k=1}^{\lfloor \frac{R}{\widetilde{R}}\rfloor }\widetilde{R}%
k(1-p_i)p_i^{k-1}+p_i^{\lfloor \frac{R}{\widetilde{R}}\rfloor }\Bigg\}}{%
\widetilde{R}} \\
\mathbb{E}[\alpha _{i}(\mathcal{K}_i)]\rightarrow & \frac{(1-p_i)\Bigg\{%
\sum_{k=1}^{\lfloor \frac{R}{\widetilde{R}}\rfloor }\widetilde{R}%
k(1-p_i)p_i^{k-1}+p_i^{\lfloor \frac{R}{\widetilde{R}}\rfloor }\Bigg\}}{%
\widetilde{R}}
\end{align}%
as $|\mathcal{K}_i|\rightarrow \infty $, where $p_i=\prod_{c\in C_{i},r\in
D_i}\sum_{n}p_{c,r}^{\text{AP}}(n)p_{i,c,r}^{n}$
\end{cor}

Due to the space constraint, the proof is presented in \cite{CORR}.

\section{Protocol Design}

Our protocol MVGMP extends the current IETF Internet standard for multicast
group management, IGMP \cite{IGMPRFC}, by adding the view selection feature
to the protocol, while each client selects one view or a set of views
according to the analytical results in Section \ref{sec: analysis}. IGMP is
a receiver-oriented protocol, where each user periodically and actively
updates its joining multicast groups to the designated router (i.e., the AP
in this paper). Due to the space constraint, this section only summarizes
the behavior of our protocol, and detailed operation can be founded in \cite%
{CORR}.

For a multi-view 3D video, each view is delivered to a multicast group since
users can receive different sets of views. The AP maintains a table, named
\textit{ViewTable}, for each video. The table specifies the current
multicast views and the corresponding bit-rates and channels for each view%
\footnote{%
Note that each view is allowed to be transmitted multiple times in differnt
channels and rates if necessary, as described in Section \ref{sec: analysis}.%
}, and each multicast view is associated with a multicast address and a set
of users that choose to receive the view. ViewTable is periodically
broadcasted to all users in the WiFi cell. MVGMP includes the following
control messages. 1) Join: A Join message contains the address of a new user
and the corresponding requested view(s), which can be the subscribed views,
or the left and right views to synthesize the subscribed view. An existing
user also exploits this message to update its requested views. 2) Leave: A
Leave message includes the address of a leaving user and the views that are
no longer necessary to be received. An existing user can also exploit this
message to stop receiving a view. Following the design rationale of IGMP,
MVGMP is also a soft-state protocol, which implies that each user is
required to periodically send the Join message to refresh its chosen views,
so that unexpected connection drops will not create dangling states in
ViewTable.

\textbf{Join. }When a new member decides to join a 3D video multicast
transmission, it first acquires the current ViewTable from the AP.
Afterward, the user identifies the views to be received according to Theorem
1. Specifically, the client first examines whether ViewTable has included
the subscribed view. If ViewTable does not include the subscribed view, or
if the view loss probability for the subscribed view in the corresponding
channel and bit-rate exceeds the threshold, the user adds a left view and a
right view that lead to the maximal decrement on the view failure
probability. The above process is repeated until the view failure
probability does not exceed the threshold.

When a multi-view 3D video starts, usually the current multicast views in
ViewTable are not sufficient for a new user. In other words, when the view
failure probability still exceeds the threshold after the user selects all
transmitted left and right views within the range $R$ in ViewTable, the user
needs to add the subscribed view to ViewTable with the most suitable channel
and bit-rate to reduce the view failure probability. Also, the left and
right views are required to be chosen again to avoid receiving too many
views. After choosing the views to be received, a Join message is sent to
the AP. The message contains the views, which the user chooses to receive,
and the AP adds the user to the ViewTable accordingly. To avoid receiving
too many views, the client can restrict the maximum number of left and right
views that are allowed to be received and exploited for DIBR.

\textbf{Leave and View Re-organization. }On the other hand, when a user
decides to leave a 3D video multicast transmission, it multicasts a Leave
message to the AP and any other user that receives at least one identical
view $k_{i}$. Different from the Join message, the Leave message is also
delivered to other staying users in order to minimize the bandwidth
consumption, since each staying user that receives $k_{i}$ will examine if
there is a chance to switch $k_{i}$ to another view $\overline{k}_{i}$ that
is still transmitted in the network. In this case, the staying user also
sends a Leave message that includes view $k_{i}$, together with a Join
message that contains view $\overline{k}_{i}$. If a view is no longer
required by any staying users, the AP stops delivering the view. Therefore,
MVGMP can effectively reduce the number of multicast views. Due to the space
constraint, an illustrative example is presented in \cite{CORR}.

\textbf{Discussion. }Note that MVGMP can support the scenario that a user
changes the desired view, by first sending a Leave message and then a Join
message. Similarly, when a user moves and thus the channel condition
changes, it will send a Join mesage to receive additional views if the
channel condition deteriorates, or a Leave message to stop receiving some
views if the channel condition improves. Moreover, when a user handovers to
a new WiFi cell, it first sends a Leave message to the original AP and then
a Join message to the new AP. If the network connection to a user drops
suddently, the AP removes the information corresponding to the user in
ViewTable when it does not receive the Join message (for soft-state update
explained early in this section) for a period of time. Therefore, MVGMP also
supports the silent leave of a user from a WiFi cell. Moreover, our protocol
can be extended to the multi-view subscription for each client by replacing
Theorem 1 with Theorem 2. The fundamental operations of
Join/Leave/Reorganize remain the same since each view is maintained by a
separate multicast group. Due to the space constraint, an illustrative
example of MVGMP is presented in \cite{CORR}.

\begin{figure*}[t]
\begin{minipage}[b]{2.2 in}
\includegraphics[width=2in, angle=270]{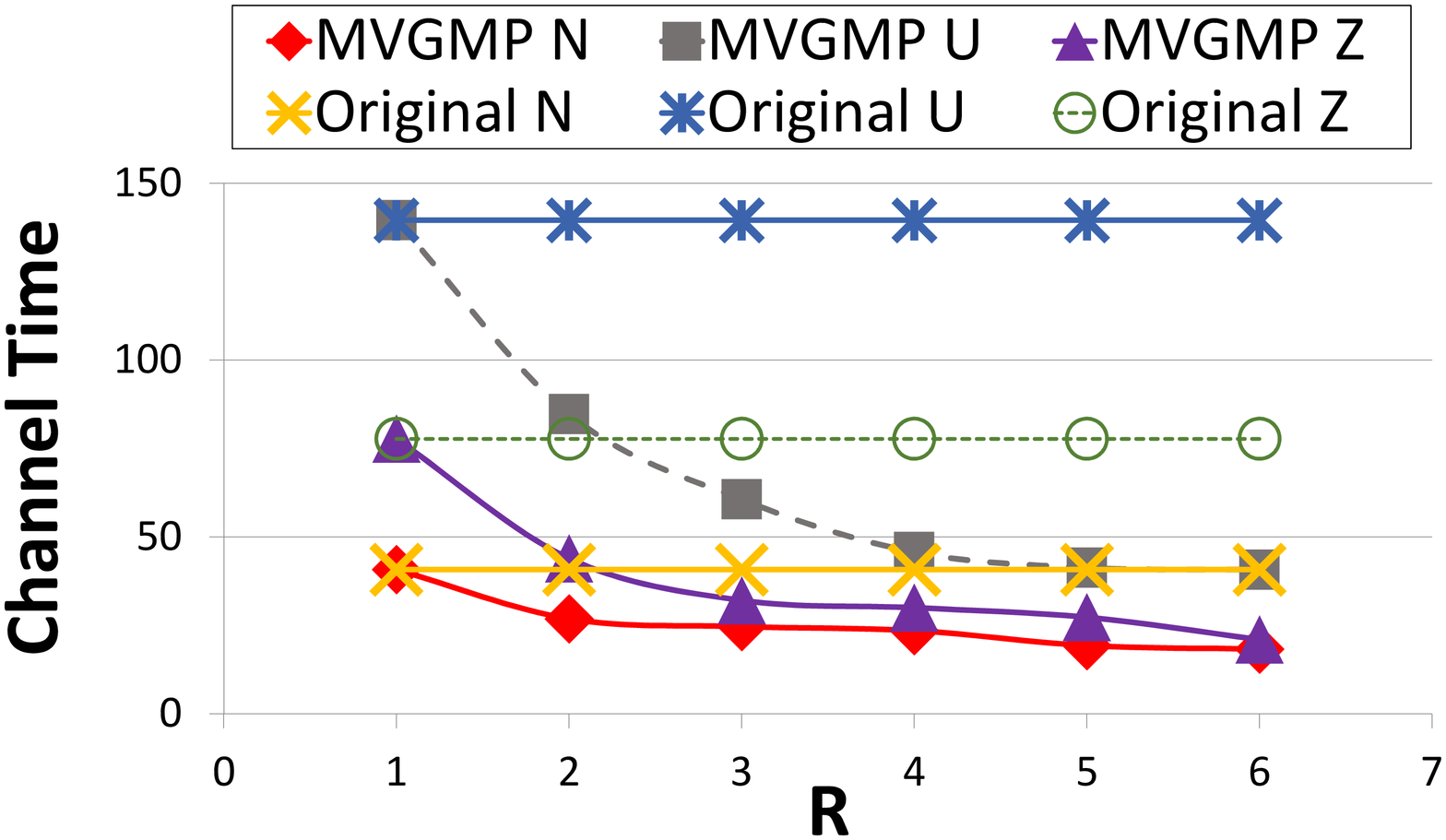}
\vspace{-1.9cm}
\caption{Synthesis Range}
\end{minipage}
\begin{minipage}[b]{2.2 in}
\includegraphics[width=2in, angle=270]{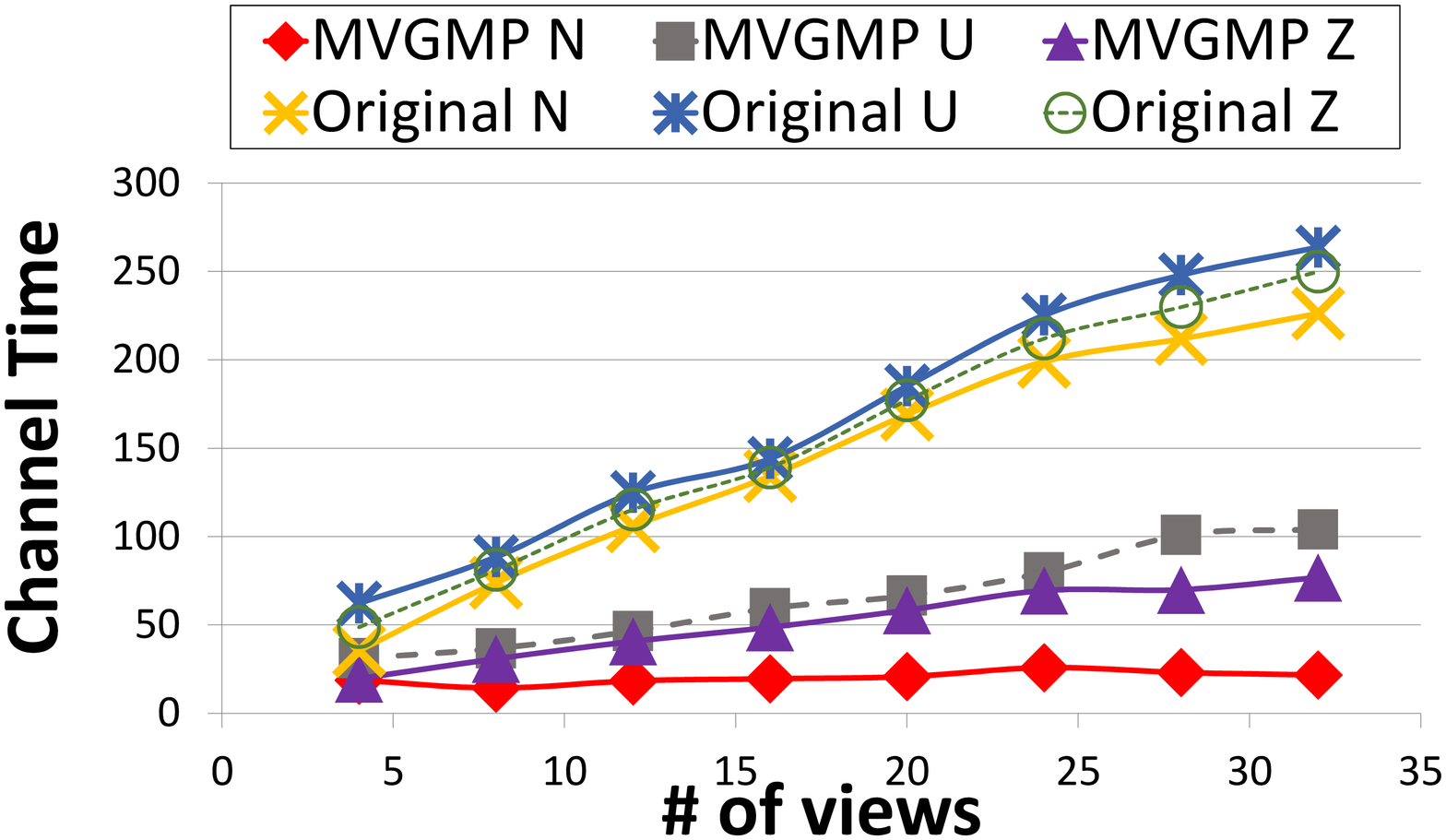}
\vspace{-1.9cm}
\caption{Number of Views in a Video}
\end{minipage}
\begin{minipage}[b]{2.2 in}
\hspace{0.2cm}\includegraphics[width=2in, angle=270]{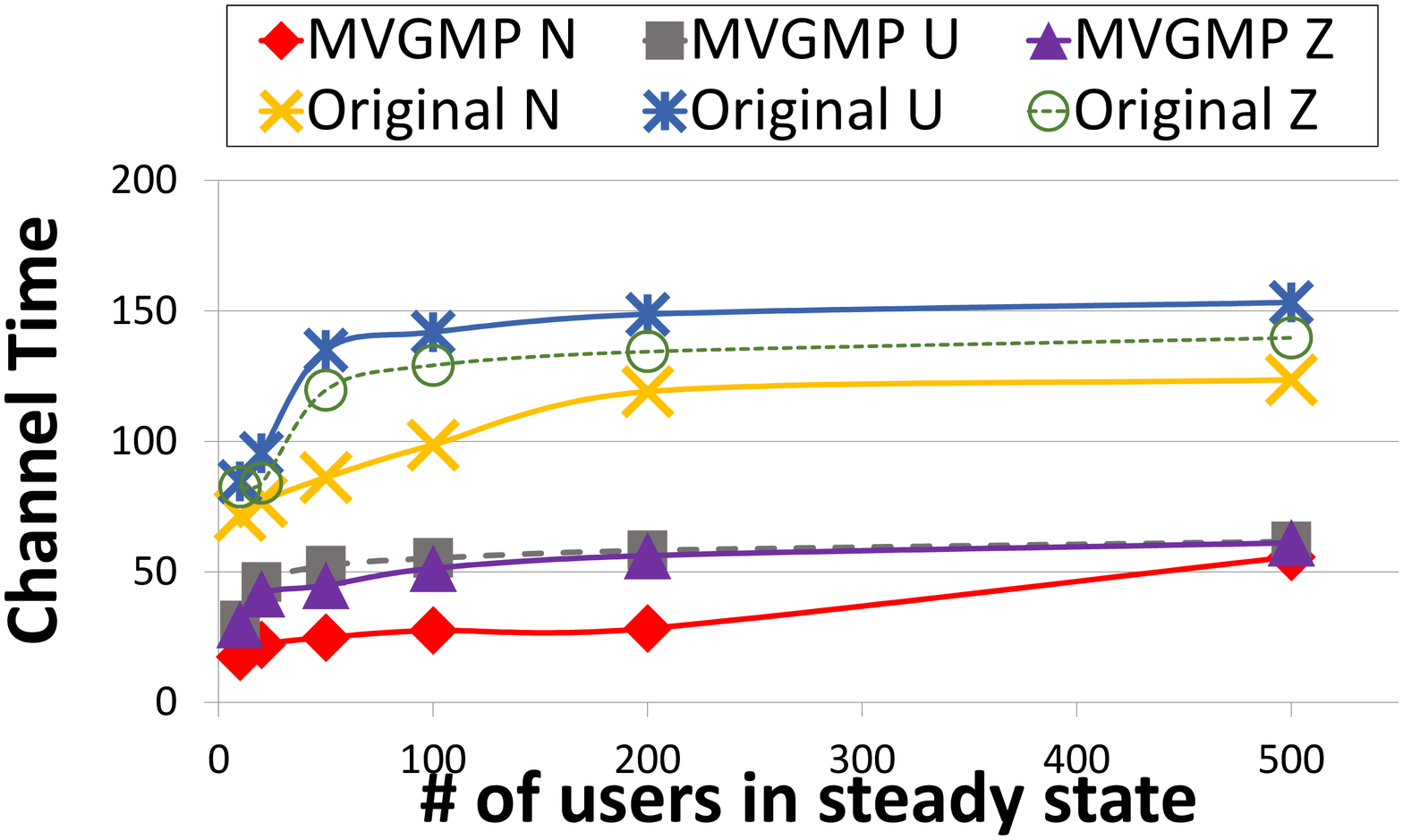}
\vspace{-1.9cm}
\caption{Number of Users}
\end{minipage}
\begin{minipage}[b]{2.2 in}
\includegraphics[width=2.2in, angle=270]{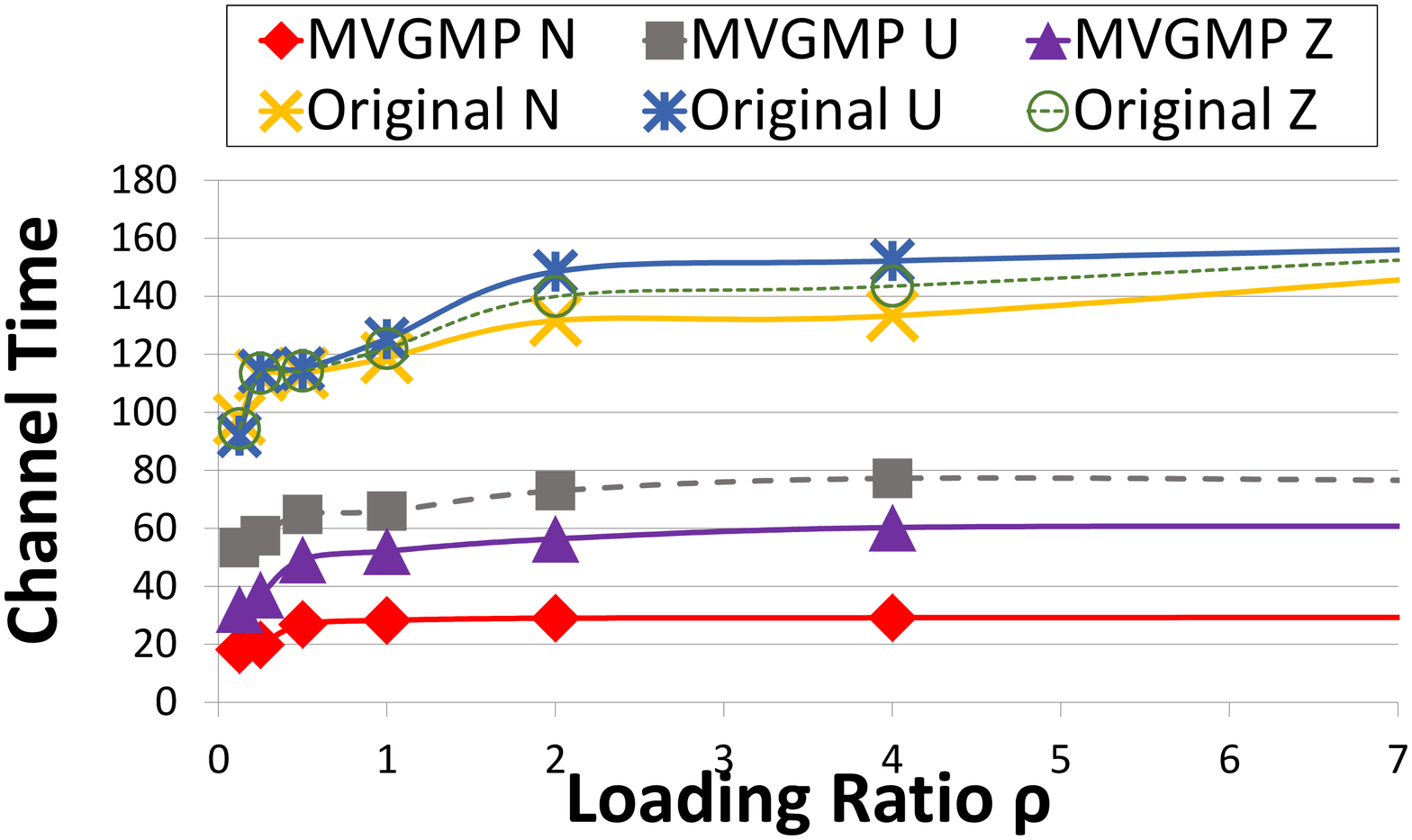}
\vspace{-2.1cm}
\caption{Network Load}
\end{minipage}
\begin{minipage}[b]{2.3 in}
\includegraphics[width=2.2in, angle=270]{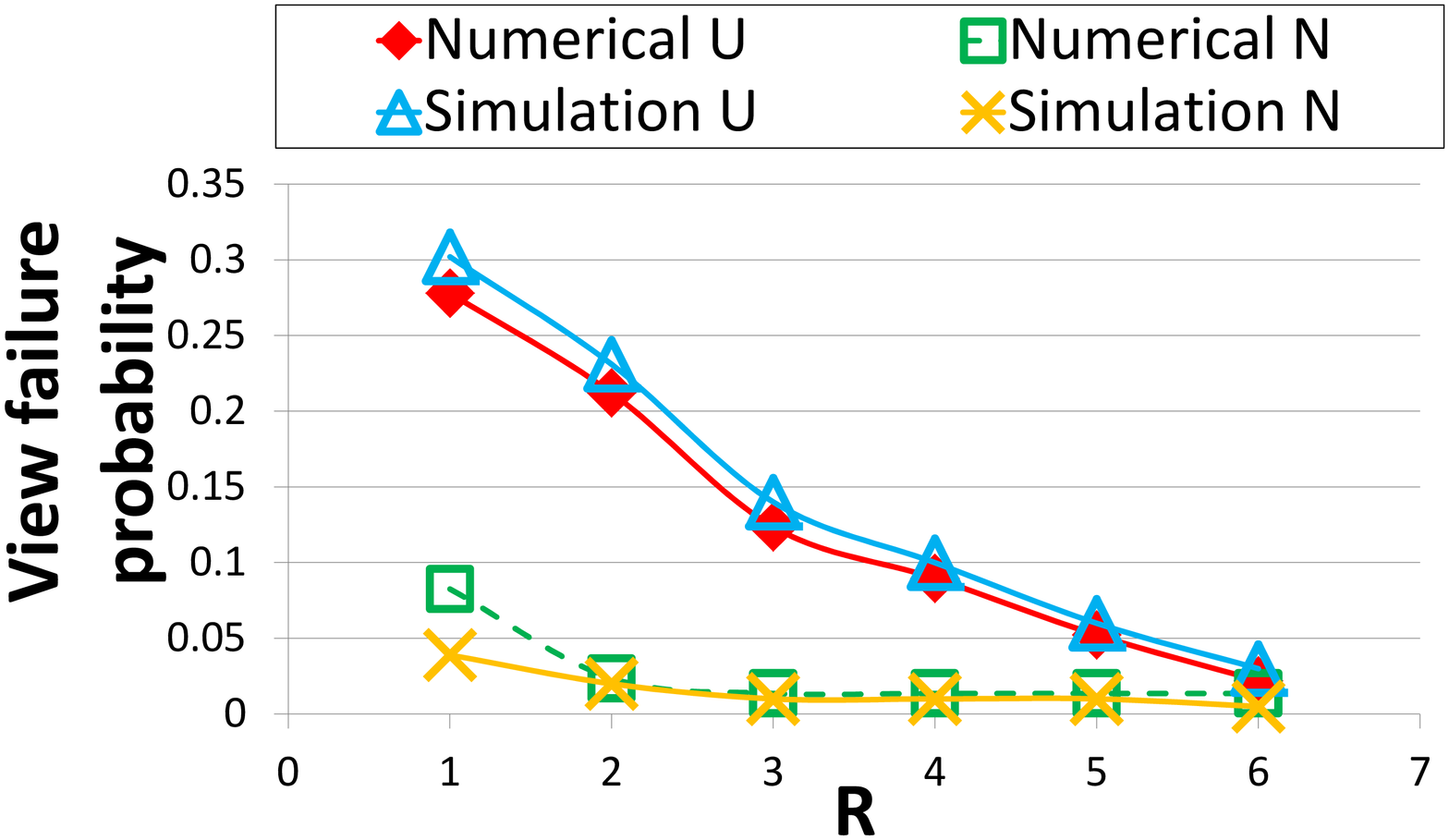}
\vspace{-2.1cm}
\caption{View Failure Probability}
\end{minipage}
\begin{minipage}[b]{2.2 in}
\hspace{0.5cm}\includegraphics[width=2.2in, angle=270]{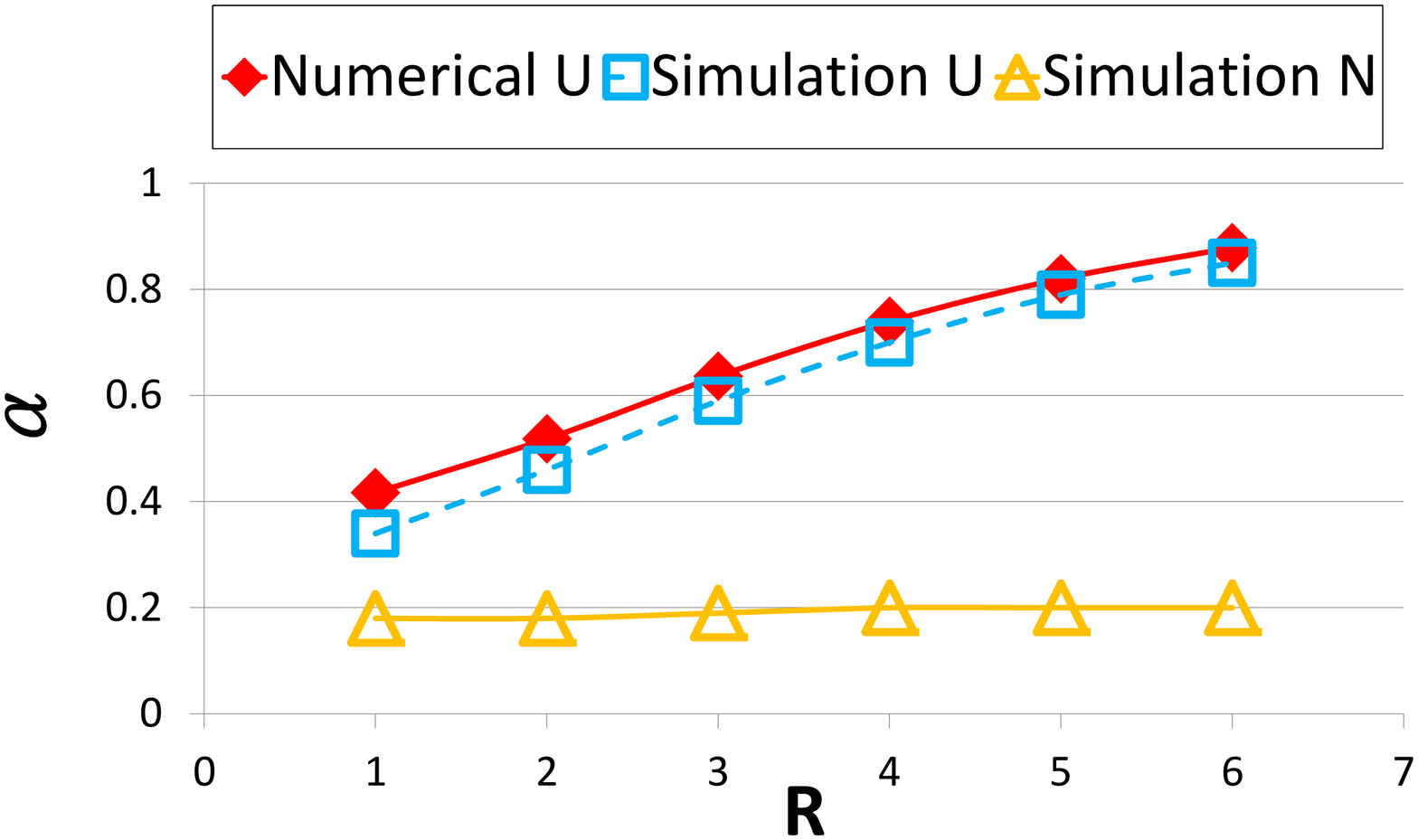}
\vspace{-2.1cm}
\caption{Ratio of Successfully Received Views}
\end{minipage}
\vspace{-10pt}
\end{figure*}

\section{Simulation Results}

In the following, we first describe the simulation setting and then compare
MVGMP with the current multicast scheme.

\subsection{Simulation Setup}

In this section, we evaluate the channel time of MVGMP in a series of
scenarios in NS3 with 802.11n program package. The channel time of a
mutlicast scheme is the average time consumption of a frame. To the best
knowledge, there is no related work on channel time minimization for
multi-view 3D video multicast in WiFi networks. Thus, we compare MVGMP with
the original WiFi multicast scheme, in which all desired views are
multicasted to the users. We adopt the setting of a real multi-view 3D
dataset, Book Arrival \cite{Meet2008} with 16 views, i.e., $|V|=16$. Each
user randomly chooses one preferred view from three preference
distributions: Uniform, Zipf, and Normal distributions. There is no
specifically hot view in Uniform distribution. By contrast, Zipf
distribution, $f(k;s;N)=(\frac{1}{k^{s}})/\sum_{n=1}{N}(\frac{1}{n^{s}})$,
differentiates the desired views, where $k$ is the preference rank of a
view, $s$ is the the exponent characterizing the distribution, and $N$ is
the number of views. The views with smaller ranks are major views and thus
more inclined to be requested. We set $s=2$ and $N=16$ in this paper. In
Normal distribution, the mean is set as $0.5$, and the variance is set as 1
throughout this paper.

We simulate a dynamic environment with $50$ client users locating randomly
in the range of an AP. After each frame, there will be an arrival and
departure of a user with probabilities $\lambda $ and $\mu $, respectively.
In addition, a user will change the desired view with probability $\eta $.
The default probabilities are $\lambda =0.2$, $\mu =0.3$, $\eta =0.4$. TABLE
II summarizes the simulation setting consisting of an 802.11n WiFi network
with 40MHz channel bandwidth and two orthogonal channels. In the following,
we first compare the performance of MVGMP with the current WiFi multicast
scheme in different scenarios and then compare the analytical and simulation
results.

\begin{table}[t]
\caption{Simulation settings.}
\label{table1}\vspace{-8pt}
\par
\begin{center}
\begin{tabular}{|l|l|}
\hline
\textbf{Parameter} & \textbf{Value} \\ \hline
Carrier Frequency & 5.0 GHz \\ \hline
The unit of Channel Time & $10^{-3}ms$ \\ \hline
Channel Bandwidth & 40MHz \\ \hline
AP Tx Power & 16.dBm \\ \hline
OFDM Data Symbols & 7 \\ \hline
Subcarriers & 108 \\ \hline
View Size (per 3D video) & 64kbits \\ \hline
Number of Orthogonal Channels & 2 \\ \hline
Data Rates & 8 different values defined \\
& in 802.11n spec. \cite{Stand2012} \\ \hline
\end{tabular}
\vspace{-12pt}
\end{center}
\end{table}

\subsection{Scenario: Synthesized Range}

Fig. 1 evaluates MVGMP with different settings of $R$. As expected, the
channel time is efficiently reduced as $R$ increases. Nevertheless, it is
not necessary to set a large $R$ because the improvement becomes marginal as
$R$ exceeds 3. Therefore, the result indicates that a small $R$ (i.e.,
limited quality degradation) is sufficient to effectively reduce the channel
time in WiFi.

\subsection{Scenario: Number of Views}

Fig. 2 explores the impact on the numbers of views in a video. The channel
time in both schemes increases when the video includes more views, because
more views are necessary to be transmitted. The result manifests that MVGMP
consistently outperforms the original WiFi multicast scheme with different
numbers of views in a video.

\subsection{Scenario: Number of Users in Steady State}

Fig. 3 evaluates the channel time with different numbers of users in the
steady state. We set $\lambda =\mu =0.25$, so that the expected number of
users in the network remains the same. The channel time grows as the number
of users increases. Nevertheless, the increment becomes marginal since most
views will appear in \textit{ViewTable}, and thus more users will subscribe
the same views in the video.

\subsection{Scenario: Utilization Factor}

Fig. 4 explores the impact of the network load. Herein, we change the
\textit{loading ratio} $\rho :=\frac{\lambda }{\mu }$, i.e., the ratio
between arrival probability $\lambda $ and departure probability $\mu $,
from $0.125$ to $8$. The results manifest that the channel time is increased
for both multicast schemes. Nevertheless, MVGMP effectively reduces at least
$40\%$ of channel time for all the three distributions.

\subsection{Impact of User Preferences}

From Fig. 1 to Fig. 4. the results manifest that Uniform distribution
requires the most channel time compared with Normal distribution. This is
because in Normal distribution users prefers a few central front views and
thus has a large probability of being synthesized with two views in range $R$%
.

\subsection{Analystical Result}

Fig. 5 and Fig. 6 compare the simulation results from NS3 and the analytical
results of Theorem 1 and Theorem 2, where each client subscribes all views
in Fig. 6. The results show that the discrepancy among the simulation and
analysis is very small. Most importantly, $\alpha $ increases for a larger $R
$ since each user can synthesize and acquire a desired view from more
candidate right and left views when the view is lost.

\section{Conclusions}

With the emergence of naked-eye mobile devices, this paper proposes to
incorporate DIBR for multi-view 3D video multicast in WiFi networks. We
first investigate the merits of view protection via DIBR and show that the
view failure probability is much smaller than the view loss probability,
while multi-view subscription for each client is also studied. Afterward, we
propose Multi-View Group Management Protocol (MVGMP) to handle the dynamic
join and leave for a 3D video stream and the change of the desired view for
a client. Simulation results manifest that our protocol effectively reduces
the bandwidth consumption and increases the probability for each client to
successfully playback the desired view in a multi-view 3D video.
\section{CoRR}
To investigate the case where user subscribes a consecutive sequence of views, we adopt the following setting. User subscribes views according to a Zipf distribution, which means the $k$th view is subscribed with probability $\frac{c}{(k\textrm{ mod }m)^s}$ independently to other views. Figure$7$ depicts this scenario using $m=5$ as an example.
\begin{figure}[t]
\includegraphics[width=3in, angle=270]{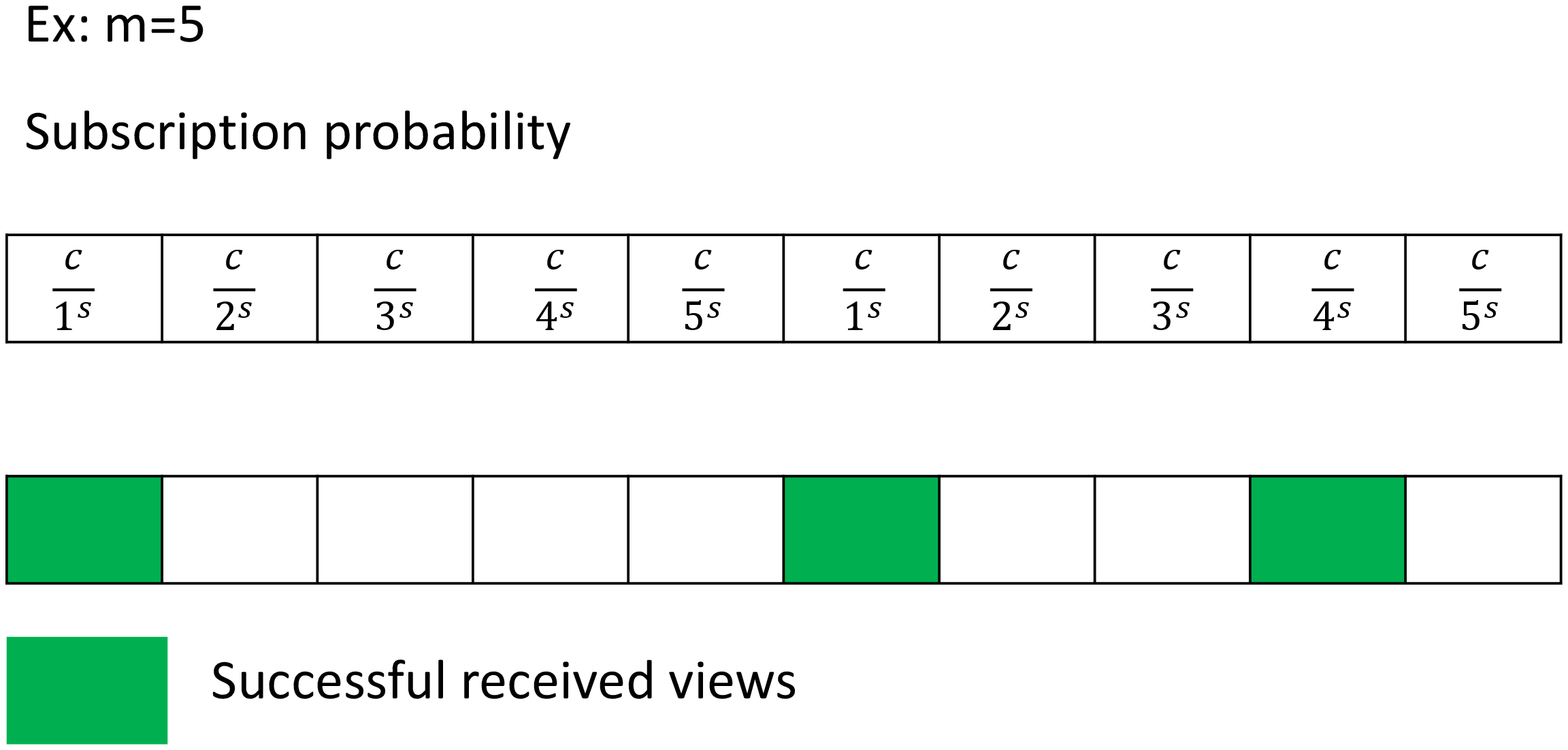}
\vspace{-3cm}
\caption{Example of consecutive view subscription scenario}
\end{figure}
Following theorem serves as a counterpart of theorem $2$ in our main article.
\begin{thm}
In the consecutive view subscription scenario as described above, the ratio $\widetilde{\alpha}$ of expected number of views that can be received or synthesized to the number of total subscribed views tends to
\begin{align}
p\frac{\sum_{j=1}^m\sum_{x=1}^R\Bigg[\Bigg(\sum_{l=1}^{m-j}\frac{c}{(j+l)^s}+\Bigg(\sum_{t=1}^m\frac{c}{t^s}\Bigg)\,
\frac{x-(m-j)}{m}\!
+\sum_{l=1}^{[x-(m-j)] \textrm{mod }m}\frac{c}{l^s}\Bigg)p(1-p)^{x-1}\Bigg]}{\sum_{l=1}^{m}\frac{c}{l^s}}
\end{align}
as $|\mathcal{K}_i|\rightarrow\infty$, where $p=1-\prod_{c\in C_{i},r\in
D_i}\sum_{n}p_{c,r}^{\text{AP}}(n)p_{i,c,r}^{n}$
\end{thm}

\textbf{Proof:} We follow a similar arguments in our main article, which derives the theorem by reward theory. This time, however, we should use a generalized reward process, the Markov reward process. Let $T_n$ denote the index of the n-th successfully received view, and $G_n$ denote the state of the embedded Markov chain, which represents the "position" of the n-th renewal cycle. An example of this definition is represented in figure. 7, in which the states of the first, second and the third cycles are $1,1,4$ respectively.

The transition probability of $G_n$ is
\begin{numcases}{p_{ij}}
   \frac{p(1-p)^{j-i-1}}{1-(1-p)^m}, & $1\leq i<j\leq m$ \nonumber\\
   \frac{p(1-p)^{m-i+j-1}}{1-(1-p)^m}, &  $1\leq j\leq i \leq m$
\end{numcases}%
since, for example $1\leq i<j\leq m$, the position change from $i$ to $j$ occurs if and only
if there are $j-i$ plus a multiple of $m$ views between the nearest two successfully received views, which means
\begin{align}
p_{ij}=p(1-p)^{j-i-1}+p(1-p)^{j-i-1+m}+p(1-p)^{j-i-1+2m}+\cdots\nonumber
\end{align}
The $\{(G_n,T_n),n=1,2,3,\dots\}$ so defined is then a Markov renewal process.

If we define the reward function of the process as
\begin{numcases}{\rho(j,x)=}
   \sum_{l=1}^x\mathbf{1(\textrm{view in the $l$ position has been subscribed})}, & $x \leq R$ \nonumber\\
   0, &  $x> R$
\end{numcases}%
then
\begin{align}
Z_{\rho}=\sum_{n:T_{n+1}<t}\rho(G_n,T_{n+1}-T_n)+\rho(G(t),X(t))
\end{align}
is a Markov reward process, where $X(t)$ is the age process and $G(t)$ be the semi-Markov process associated with our interested Markob renewal process $\{(G_n,T_n),n=1,2,3,\dots\}$.
The process so defined as the following desired property:
The process just defined has a direct relation to our desired quantity $\widetilde{\alpha}$, which is
\begin{align}
\widetilde{\alpha}=\frac{EZ_{\rho}}{S_t}\nonumber
\end{align}
where $S_t$ is the number of views subscribed by the user.
We now intend to apply the theorem 4.1 in \cite{soltani1998} to the right hand side of the above equation. In the following, we will use the same notations as in the article just mentioned.
\begin{align}
h(j)&=\sum_{x=1}^{\infty}\rho(j,x)\sum_{j=1,2,\dots}P(G_{n+1}=j,T_{n+1}-T_n= x|G_n=i)\nonumber\\
&=\sum_{x=1}^{\infty}\rho(j,x)p(1-p)^{x-1}\nonumber\\
&=\sum_{x=1}^R\Bigg[\Bigg(\sum_{l=1}^{m-j}\frac{c}{(j+l)^s}+\Bigg(\sum_{t=1}^m\frac{c}{t^s}\Bigg)\,
\frac{x-(m-j)}{m}\!\nonumber\\
&+\sum_{l=1}^{[x-(m-j)] \textrm{mod }m}\frac{c}{l^s}\Bigg)p(1-p)^{x-1}\Bigg]\nonumber\\
\end{align}
Observe that the steady state of the chain $G_n$ is uniform distribution, which means
\begin{align}
\pi_i=\frac{1}{m}
\end{align}
Now apply theorem 4.1 in \cite{soltani1998}, we have
\begin{align}
\mathbb{E}Z_{\rho}(t)=pt\sum_{j=1,2,\dots}\pi_jh(j)+o(t)\nonumber
\end{align}
Hence,
\begin{align}
&\frac{\mathbb{E}Z_{\rho}(t)}{S_t}\rightarrow mp\frac{\sum_{j=1,2,\dots}\pi_jh(j)}{\sum_{l=1}^{m}\frac{c}{l^s}}\nonumber\\
&=p\frac{\sum_{j=1}^m\sum_{x=1}^R\Bigg[\Bigg(\sum_{l=1}^{m-j}\frac{c}{(j+l)^s}+\Bigg(\sum_{t=1}^m\frac{c}{t^s}\Bigg)\,
\frac{x-(m-j)}{m}\!
+\sum_{l=1}^{[x-(m-j)] \textrm{mod }m}\frac{c}{l^s}\Bigg)p(1-p)^{x-1}\Bigg]}{\sum_{l=1}^{m}\frac{c}{l^s}}\nonumber
\end{align}

\linespread{0.90}
\bibliographystyle{IEEEtran}
\bibliography{IEEEabrv,reference}

\end{document}